\documentclass{article}

\PassOptionsToPackage{numbers, compress}{natbib}

\usepackage[preprint]{neurips_2024}
\usepackage{graphicx} 
\usepackage{amsmath}

\usepackage{xspace}
\usepackage[utf8]{inputenc} 
\usepackage[T1]{fontenc}    
\usepackage{hyperref}       
\usepackage{url}            
\usepackage{booktabs}       
\usepackage{amsfonts}       
\usepackage{nicefrac}       
\usepackage{microtype}      
\usepackage{xcolor}         
\newcommand{\method}{GenRank\xspace}
\usepackage{wrapfig}
\usepackage{footmisc}

\title{Towards Large-scale Generative Ranking}

\author{%
  \begin{tabular}[t]{c}
Yanhua Huang, Yuqi Chen, Xiong Cao, Rui Yang, Mingliang Qi, Yinghao Zhu \\ Qingchang Han, Yaowei Liu, Zhaoyu Liu, Xuefeng Yao, Yuting Jia, Leilei Ma, Yinqi Zhang \\ Taoyu Zhu, Liujie Zhang, Lei Chen, Weihang Chen, Min Zhu, Ruiwen Xu, Lei Zhang\end{tabular} \\
  Xiaohongshu Inc.\\
  Shanghai, China \\
}

\begin{document}

\maketitle

\begin{abstract}
Generative recommendation has recently emerged as a promising paradigm in information retrieval. 
However, generative ranking systems are still understudied, particularly with respect to their effectiveness and feasibility in large-scale industrial settings.
This paper investigates this topic at the ranking stage of Xiaohongshu's Explore Feed, a recommender system that serves hundreds of millions of users.
Specifically, we first examine how generative ranking outperforms current industrial recommenders. Through theoretical and empirical analyses, we find that the primary improvement in effectiveness stems from the generative architecture, rather than the training paradigm. To facilitate efficient deployment of generative ranking, we introduce \method, a novel generative architecture for ranking. We validate the effectiveness and efficiency of our solution through online A/B experiments. The results show that \method achieves significant improvements in user satisfaction with nearly equivalent computational resources compared to the existing production system.

\end{abstract}

\section{Introduction}

\begin{figure}
    \centering
    \includegraphics[width=0.8\linewidth]{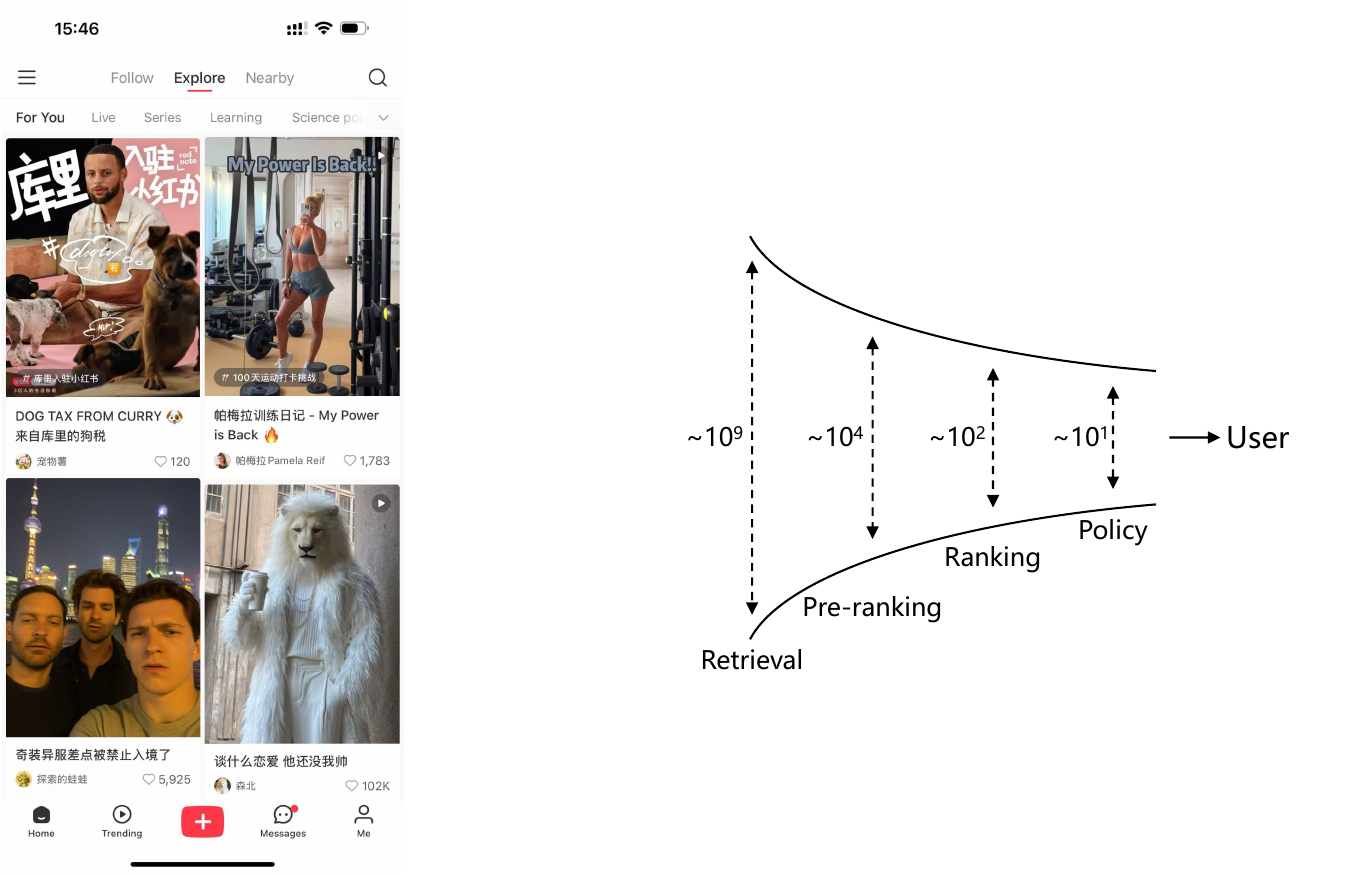}
    \caption{Left: A screenshot of Xiaohongshu's Explore Feed product. Right: Illustration of cascade pipeline for industrial recommender systems, where each stage needs to process a large number of items.}
    \label{fig:pipeline}
\end{figure}

Recommender systems are essential components of social media platforms, enabling users to browse and engage with personalized item suggestions~\citep{cheng2016wide,chang2023pepnet,huang2021sliding,zhai2024actions}. 
To balance efficiency and effectiveness, industrial recommender
systems typically employ a cascade pipeline~\citep{huang2021sliding,zhai2024actions}, consisting of four stages, as shown in Figure~\ref{fig:pipeline} (right). The retrieval stage initially selects tens of thousands of candidates from billions of items. This is followed by the pre-ranking stage, which performs coarse matching to narrow the candidate set to hundreds. The ranking stage then makes precise predictions for each candidate. Finally, the policy stage re-ranks dozens of candidates based on sequential information and commercial considerations to produce the final recommendation.

In modern recommender systems, the ranking stage generally follows the MLP \& Embedding paradigm~\citep{cheng2016wide}, where sequential modeling has achieved notable success in capturing user interests~\citep{kang2018self,zhou2018deep,sun2019bert4rec,pi2020search}. The advent of generative recommendation has further enhanced the capabilities of sequential approaches. 
Unlike traditional approaches, generative recommendations formulate the recommendation problem as a sequence generation task~\citep{rajput2023recommender,zhai2024actions}, directly predicting target behaviors from historical user actions. ~\citet{rajput2023recommender} propose achieving generative retrieval by quantizing items with hierarchical semantic IDs. ~\citet{yang2025sparse} further introduce a coarse-to-fine generation process to address information loss caused by quantization. Despite their novelty, generative recommenders for ranking tasks remain understudied, especially in large-scale industrial contexts.

This paper investigates generative ranking systems in large-scale industrial scenarios. In particular, we first analyze potential sources of effectiveness in generative recommendations and then conduct experiments to validate our hypotheses using existing generative recommenders~\citep{zhai2024actions}. Experimental results demonstrate that the generative architecture is critical to achieving strong performance. However, current generative architectures tend to be inefficient, particularly in large-scale settings. To address this issue, we propose a novel architecture, \method, to meet the requirements of large-scale training and inference. Online A/B experiments on the Explore Feed in Xiaohongshu\footnote{Also known as RedNote: https://www.xiaohongshu.com/explore.} (Figure~\ref{fig:pipeline} (left)), a recommender system serving hundreds of millions of users, demonstrate the effectiveness and efficiency of our proposed solution.

The main contributions of this paper are summarized as follows:
\begin{itemize}
    \item[1.] We identify and analyze the sources of effectiveness in generative recommendation, highlighting the critical role of generative architecture in overall performance.
    \item[2.] We propose an efficient generative architecture, specifically designed for industrial scenarios, which includes an action-oriented sequence organization approach and novel strategies for position and time biases.
    \item[3.] We conduct large-scale online A/B testing to demonstrate the effectiveness and feasibility of generative ranking in industrial recommender systems.
\end{itemize}

\section{Problem Setup}
This paper studies generative recommendations at the ranking stage. Here, the recommender system is required to make predictions for a set of predefined tasks, such as predicting the click-through probability or the expected duration a user spends when presented with a candidate item. To construct our dataset for offline experiments, we collect hundreds of billions of item exposure logs from Xiaohongshu's Explore Feed over 15 days. There are three types of input features: 
\begin{itemize}
    \item Categorical features: user ID, item ID, user historical behaviors, hashtags, etc.
    \item Numerical features: user age, item publish time, number of author fans, etc.
    \item Frozen embeddings: multi-modal item embeddings, graph-based author embeddings, etc.
\end{itemize}
Following prior work~\citep{cheng2016wide,zhang2024wukong}, numerical features are discretized into categorical features using predefined boundaries, while categorical features are transformed into dense embeddings via embedding tables. The frozen embeddings provided by pre-trained models are treated as side information, offering prior knowledge relevant to the features they are associated with.
We use the area under the ROC curve (AUC) as the offline evaluation metric. Notably, an absolute increase of 0.0010 in AUC for the main tasks is considered significant in our settings, as it typically yields a 0.5\% improvement in topline metrics for hundreds of millions of users online. 

\section{Source of Effectiveness in Generative Recommendation}
There have been numerous works on generative recommendation~\citep{zhai2024actions,yang2025sparse,ren2024non,rajput2023recommender}. However, the effectiveness of generative ranking, particularly in large-scale industrial settings, has been underexplored. To better understand the factors contributing to the effectiveness of generative ranking, we conduct experiments from two perspectives:
\begin{itemize}
    \item[1.] The generative recommendation paradigm diverges from conventional approaches through distinct underlying mechanisms. We are especially interested in identifying mechanisms where minor modifications cause major performance drops, as these may be critical to the success of generative ranking methods.
    \item[2.] The current ranking paradigm integrates several well-established modules, such as SIM~\citep{pi2020search} and content embeddings~\citep{zhang2024notellm,huang2021sliding}. 
    We examine key modules that show marked performance differences in generative settings, providing valuable insights for future studies.
\end{itemize}

Specifically, we choose HSTU~\citep{zhai2024actions} as the baseline model to present our findings from the above perspectives. By default, the number of blocks is 3, the number of attention heads is 8, and the hidden dimension is 768. Each user sequence has a maximum length of 480, including both historical behaviors and candidate items. We use the mixed precision training strategy on NVIDIA H20 GPUs. 

\subsection{Key Mechanisms in Generative Paradigm}
In contrast to conventional paradigms that learn sophisticated feature interactions from historical behaviors, generative recommendation reformulates the ranking as a sequential transduction task~\citep{zhai2024actions}. In this context, the generative ranking differs significantly in two ways: the manner of sequential interactions and the organization of training samples.

\textit{The manner of sequential interactions} in generative ranking is auto-regressive. Note that HSTU computes the loss only at positions corresponding to candidate items, as shown in Figure~\ref{fig:model}(a). This approach can be regarded as supervised fine-tuning, where the user information and candidate items serve as the input prompt. One reason modern LLMs employ the auto-regressive manner during supervised fine-tuning is to retain abilities acquired during pre-training. However, generative ranking does not involve a pre-training stage. This raises the question: Is the auto-regressive manner truly necessary for generative ranking?

To investigate this question, we conduct two sets of experiments. In the first set, we compute the loss at positions corresponding to historical behaviors. We observe a decrease in AUC of more than 0.0100, even when only a small number of historical positions are included. We attribute this to the one-epoch issue described in~\citet{zhang2022towards}, in which the model learns incorrect patterns from sparse features. In the second set, we replace the causal mask with a fully visible mask at historical positions. This modification is analogous to the T5 model~\citep{raffel2020exploring}, in which the attention mask maximizes feature interaction across the prompt. However, this change causes an AUC drop of more than 0.0015, and the decline grows more significant with larger model sizes. These results support the conclusion that the auto-regressive manner is critical to the effectiveness of generative ranking.

\textit{The organization of training samples} in conventional paradigms is typically point-wise; that is, each training sample corresponds to an item exposure log. In contrast, generative ranking groups the temporally adjacent behaviors of a user into a single training sample. We hypothesize two possible benefits of this organization. First, since two exposure logs from the same request overlap significantly in features (particularly user features), processing them in the same batch improves gradient estimation stability. Second, we consider a practical viewpoint. In large-scale online distributed training, the order in which samples are processed does not strictly follow the actual temporal order, potentially causing information leakage. Under such conditions, the model can deduce the user preference for an item from historical behavior features prior to observing its exposure log during training.
The organization in generative ranking helps mitigate the risk of training later-occurring samples before earlier ones.

However, our empirical results do not strongly support these two hypotheses. Specifically, we train the generative recommender with grouped training samples but in a point-wise order to simulate conventional training. This approach results in only a slight decrease in AUC. Therefore, we conclude that the effectiveness of generative recommendations primarily stems from the architecture, rather than the way training samples are organized.

\subsection{Module Performance Comparison across Paradigms}
To compare the influence of modules between the two paradigms, we perform experiments measuring the performance gains achieved by various modules. In particular, we select four important modules that are commonly employed in industrial ranking systems: SIM~\citep{pi2020search} for sequential modeling, PPNet~\citep{chang2023pepnet} for personalized representation learning, content embeddings~\citep{huang2021sliding,zhang2024notellm,zhai2023sigmoid} for prior knowledge, and PLE~\citep{tang2020progressive} for multi-task learning. The results show that SIM, PPNet, and PLE achieve comparable improvements in both paradigms, suggesting that the generative paradigm is compatible with these modules.
Furthermore, we observe that content embeddings yield over twice the AUC improvement under the generative paradigm. We attribute this enhancement to the architectural consistency between the generative training of content embeddings and their application in downstream tasks, allowing optimal utilization of their capabilities.

We also study the impact of feature engineering, which is critical for the performance of industrial recommendations~\citep{huang2022neural}. HSTU~\citep{zhai2024actions} proposes to remove these features because generative recommenders can express statistical patterns sufficiently. Our experiments show that while the majority of features provide negligible benefits to generative architectures, certain real-time statistical features, especially window-based ones, remain remarkably effective at enhancing performance. We posit that these features offer direct signals to the model, allowing the generative architecture to learn complex patterns. Notably, the significant computational overhead associated with feature engineering limits ranking models' ability to process large candidate sets in real time. Generative architectures address this limitation through their minimal feature engineering requirements, thereby enhancing inference scalability. Furthermore, the KV cache mechanism enables generative architectures to scale more efficiently with increasing candidate set sizes~\citep{zhai2024actions}. We envision that with continued reductions in computational overhead, generative architectures could potentially unify ranking and pre-ranking stages in future systems.

\section{Efficient Generative Ranking in Industrial Scenarios}
\label{sec:deploy}

The previous section highlights the importance of architecture in generative ranking. Not only is it crucial for performance, but it also impacts the overall design of future recommendation systems. This section introduces a novel generative architecture, \method, to enable efficient training and inference on large-scale ranking tasks. \method differs from existing work in two ways: item-action organization (Section~\ref{sec:org}) and position \& time biases (Section~\ref{sec:pt}).

Table~\ref{tab:exp} summarizes our empirical results in training performance. We use HSTU~\citep{zhai2024actions} as the baseline method. Transitioning to the action-oriented organization yields a speed-up of 78.7\%, while adopting the proposed position \& time biases yields a speed-up of 25.0\%. In general, \method achieves a total speed-up of 94.8\% during training, with a slight improvement in AUC on the test set.
\begin{table}[hbp]
    \centering
    \begin{tabular}{c|c|c}
    \toprule
    Variant & Speed-Up & AUC Diff \\
    \midrule
    Baseline (HSTU) & / & / \\
    \midrule
    + Action-oriented Organization & +78.7\% 
 & -0.0003\\
    + Proposed Position \& Time Biases & +25.0\% & +0.0009 \\
    \midrule
    + All (\method) & +94.8\% & +0.0006 \\
    \bottomrule
    \end{tabular}
    \vspace{0.1in}
    \caption{Ablation study of \method.}
    \label{tab:exp}
\end{table}

\subsection{Item-Action Organization}~\label{sec:org}

\begin{figure}[htbp]
    \centering
    \vspace{-0.2in}
    \includegraphics[width=\linewidth]{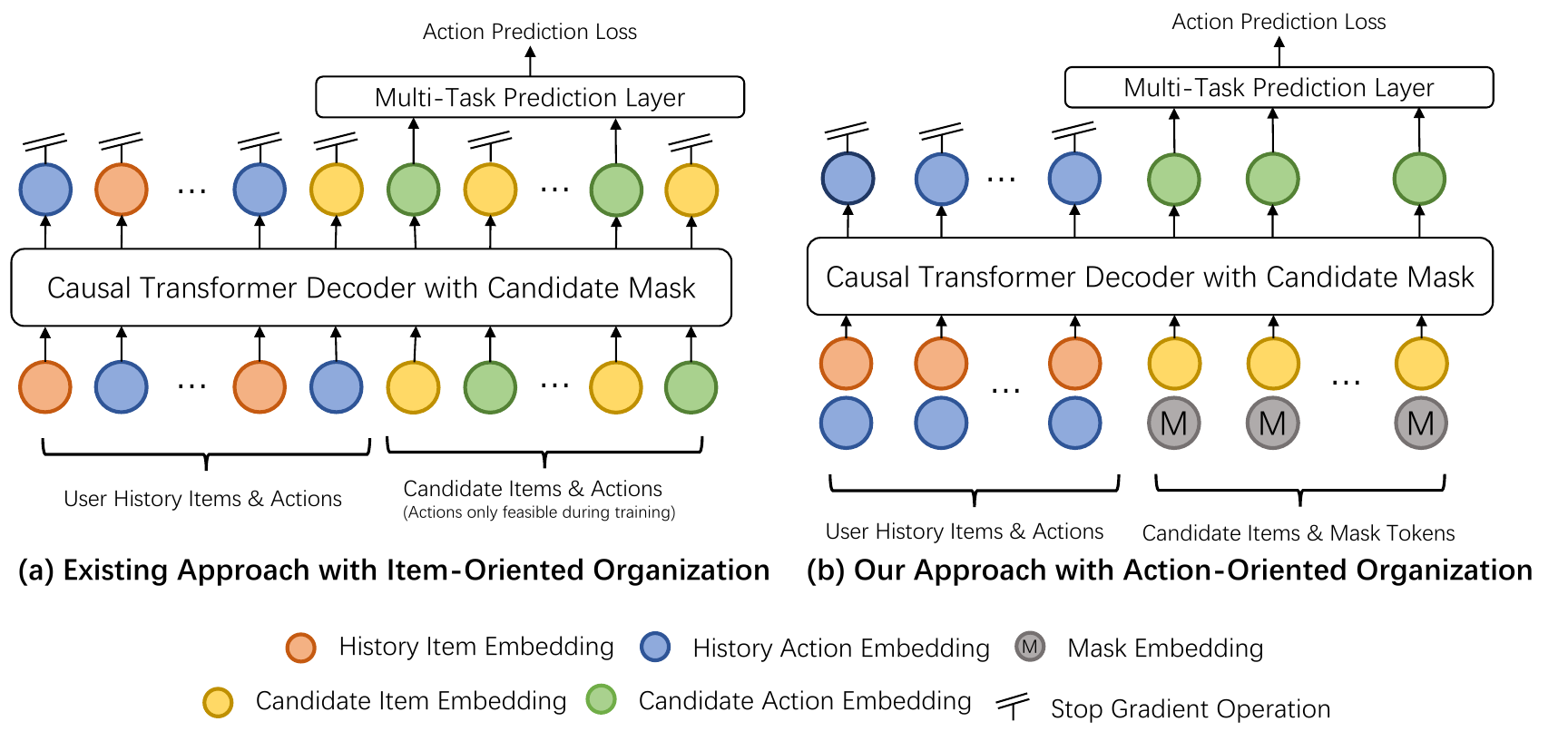}
    \caption{Model architecture of \method. Compared to existing approaches, e.g., HSTU~\citep{zhai2024actions}, which adopt an item-oriented organization, our solution adopts an action-oriented organization.}
    \label{fig:model}
\end{figure}

Conventional sequential recommendation methods typically construct models by treating individual items as fundamental units~\citep{zhou2018deep, kang2018self, sun2019bert4rec,chang2023twin,si2024twin}, an organizational framework we term the item-oriented architecture. To adapt these approaches to meet the action-aware formulation for ranking tasks, HSTU~\citep{zhai2024actions} treats action tokens as an additional modality in the sequence. As illustrated in Figure~\ref{fig:model}(a), it interleaves items and actions in a single sequence, thus enabling the model to predict either an item or an action based on the contextualized sequence.
Although such approaches can support retrieval and ranking tasks within a unified framework, they introduce substantial overhead for ranking because the sequence length is doubled. 

To address this limitation, we propose a new perspective: \textbf{we treat items as positional information and focus on iteratively predicting the actions associated with each item}, which we refer to as the action-oriented organization. In this paradigm, actions become the fundamental units in sequence generation, while items serve as contextual signals to guide the generative process, as depicted in Figure~\ref{fig:model}(b). This approach focuses on action prediction and offers significant advantages in terms of efficiency. This design halves the attention mechanism’s input sequence length, cutting attention costs by 75\% and linear projection costs by 50\%.

\begin{figure}[htbp]
    \centering
    \includegraphics[width=\linewidth]{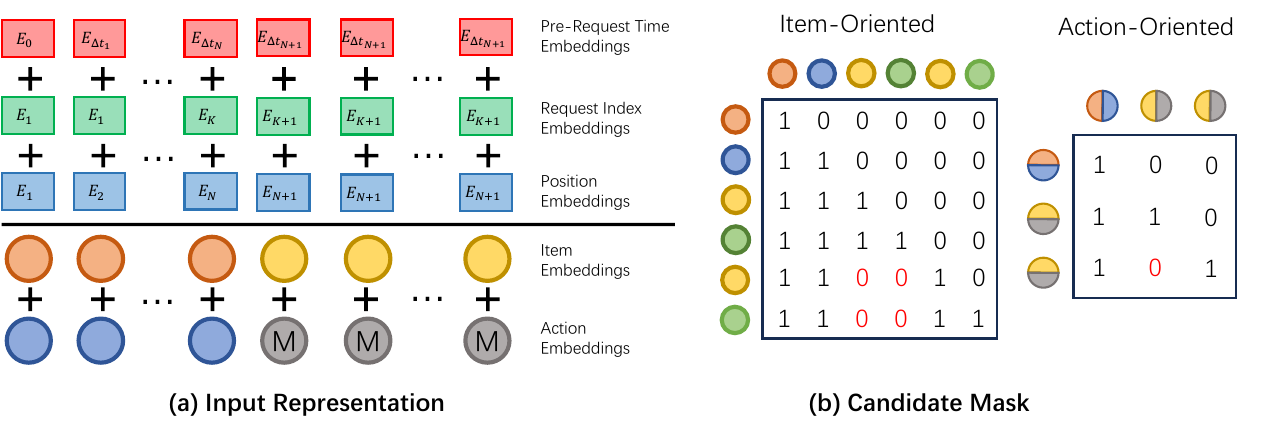}
    \caption{Illustration of input representation and candidate mask. (a) The input representation of \method contains five kinds of embeddings. (b) Candidate Mask: We depict the mask structure for a user's sequence with one historical behavior and two candidate items.} 
    \label{fig:position}
\end{figure}

Formally, we consider a list of $N$ user tokens $x_1, x_2, ..., x_N$, ordered chronologically, where $x_i \in \mathcal{X}$ (the set of items). For each item $x_i$, there is an associated action $a_i \in \mathcal{A}$ (the set of actions), which occurs at timestamp $t_i$. Thus, the sequence of actions is $a_1, a_2, ..., a_N$ and the corresponding timestamps are $t_1, t_2, ..., t_N$.
In our setup, the model learns to approximate the distribution $p(a_k|x_1, a_1, ..., x_k)$. 
To implement an action-oriented generative ranking, each input token combines both item and action embeddings, as shown in Figure~\ref{fig:position}(a). For each position in the user's history sequence, the token embedding is obtained by the sum of the item embedding and the action embedding, i.e., $e_i = \varphi(x_i)+\phi(a_i)$, where $\varphi(\cdot)$ and $\phi(\cdot)$ denote the item and action embedding modules, respectively. Our task is to predict the user's action on the next candidate item. To achieve this, the token embedding of a candidate item is given by $e_j=\varphi(x_j)+M$, where $M$ is a mask action embedding. Note that, to prevent information leakage between candidates, a candidate mask is applied, as illustrated on the right side of Figure~\ref{fig:position}(b).


\subsection{Position \& Time Biases} \label{sec:pt}

HSTU~\citep{zhai2024actions} utilizes a learnable relative attention bias to encode position and time information. Although this design is critical for performance~\citep{guo2024scaling}, it introduces a computational bottleneck: the I/O operations for attention biases scale quadratically with sequence length, incurring significant overhead as the context window grows. 
This inefficiency motivates us to design new position \& time biases that significantly reduce system cost. Specifically, we present a comprehensive design for position and time embeddings, requiring only linear I/O operations, including:
\begin{itemize}
    \item Position Embeddings: A learnable positional embedding is used to record the index of items in the user sequence, denoted as $E_{pe,i} = \Omega_{pe}(i)$. To ensure consistency between training and inference, candidate items within the same request share the same position.
    \item Request Index Embeddings: In practice, a user can interact with multiple items within a single request. We treat all items belonging to the same request as a group and define the request index embedding $E_{ri,i}$ as $\Omega_{ri}(|\{t_1, ..., t_i\}|)$, where $|{\cdot}|$ represents the cardinality.
    \item Pre-Request Time Embeddings: This embedding captures the bucketed time difference between each item and the time of the previous request, reflecting the user's activity level. Specifically, it is defined as $E_{rt,i} = \Omega_{rt}(\text{bucket}(t_i - \max_{t_j<t_i}{ t_j }))$.
\end{itemize}

The above design introduces minimal training overhead while preserving positional and temporal information. Finally, the input representation fed into the subsequent network is:
\begin{equation}
e^{(p,t)}_i=\varphi(x_i)+\phi(a_i)+E_{pe,i}+E_{ri,i}+E_{rt,i} ~.
\end{equation}

Additionally, a critical limitation of the above position and time embeddings is the lack of interaction between time and position information. To address this, we propose to employ a parameter-free bias, ALiBi~\citep{press2021train}, as the relative position \& time biases in attention mechanisms. ALiBi has two main advantages. It penalizes attention scores between distant query-key pairs, with the penalty increasing as the distance between a key action token and a query action token grows. We believe this design is more aligned with the pattern of user interest modeling. Moreover, ALiBi is parameter-free, i.e., it does not require $O(N^2)$ memory access overhead or gradient backpropagation.
By fusing ALiBi into the flash attention~\citep{xu2024enhancing}, we incur only minimal computational cost.

\section{Related Work}

\subsection{Generative Recommendation}

Generative recommendation has emerged as a promising paradigm in information retrieval~\citep{rajput2023recommender,yang2025sparse,wang2024content,wang2024eager,cai2025exploring}. Unlike traditional recommendation approaches, generative recommendation aims to directly generate recommendations from a user's historical behaviors by formulating the recommendation as a sequence generation task. TIGER~\citep{rajput2023recommender} is the first generative retrieval framework. It first acquires hierarchical IDs of items by quantizing their semantic embeddings, and then trains a sequence-to-sequence model that predicts the semantic ID of the next item. ColaRec~\citep{wang2024content} and LETTER~\citep{wang2024learnable} study the problem of enhancing collaborative signals in quantization to integrate content knowledge and collaborative interaction. COBRA~\citep{yang2025sparse} addresses the information loss from the quantization by a coarse-to-fine generation mechanism, enabling more expressive generative modeling. 
Despite their advancements, the effectiveness and feasibility of generative ranking are still underexplored in large-scale scenarios in the real world. HSTU~\citep{zhai2024actions} is the first study to investigate the generative ranking task. It introduces an interleaved organization to predict the action by treating user actions as a new modality. 
In contrast, \method views items as positional indicators and reformulates recommendation as an action-oriented generation problem. Furthermore, we systematically analyze the efficacy drivers in generative recommendation, yielding crucial insights for both understanding generative ranking paradigms and informing future architectural designs.

\subsection{Scaling Law in Recommendation System}

Scaling laws, well-established in natural language processing and computer vision~\citep{kaplan2020scaling,zhai2022scaling}, describe predictable relationships between model performance and factors such as model size, dataset size, and computational resources. In the domain of recommendation systems, similar scaling behaviors have been observed and validated across various stages of the pipeline, including retrieval~\citep{zhai2024actions,guo2024scaling,cai2025exploring} and ranking~\citep{zhai2024actions,zhang2024wukong}. Among recent advances, HSTU~\citep{zhai2024actions} emerges as a promising approach for generative recommendation. However, deploying such models in large-scale real-world scenarios necessitates careful consideration of efficiency. In this paper, we introduce an efficient generative architecture for ranking tasks while maintaining a comparable overhead to that of the current industrial recommender.

\section{Online Experiments}
To validate the effectiveness and feasibility of generative ranking in product scenarios, we conducted online experiments in Xiaohongshu's Explore Feed. All models traced back more than three months of data and were trained in an online manner. For the control group, we randomly selected 10\% of users of Xiaohongshu
and applied the production ranking model. For
the treatment group, we applied \method to a randomly selected 10\% of users. Each group contains tens of millions of users, and there is no overlap between groups.

\begin{table}[htbp]
    \centering
        \begin{tabular}{c|cccc}
        \toprule
             & Time Spent & Reads & Engagements & LT7 \\
        \midrule
            Improvement & +0.3345\% & +0.6325\% & +1.2474\% & +0.1481\% \\
        \bottomrule
        \end{tabular}
    \vspace{0.1in}
    \caption{Online A/B test result in the Explore Feed scenario.}
    \label{tab:ab}
\end{table}

From the offline metrics, the improvements in both AUC and GAUC~\citep{zhou2018deep,chang2023pepnet} for primary tasks exceed 0.0020, while the improvements for other tasks range between 0.0005 and 0.0015. From the online metrics, we select four metrics to measure the online performance: time spent, the number of reads, the number of engagements, and lifetime over 7 days (LT7). The online A/B test result averaged over a 15-day experimental period is shown in Table~\ref{tab:ab}, where \method outperforms the production ranking in all metrics. Specifically, we observe that the improvement of \method on cold-start items is particularly significant. We believe this improvement stems from \method's enhanced ability to leverage the world knowledge from content embeddings.

In terms of overhead, \method and the production ranking model require a comparable amount of overall resources. Specifically, \method incurs higher training costs but lower inference and storage costs. Furthermore, \method demonstrates a significant improvement in P99 response time, outperforming the production ranking model by over 25\%. This highlights the potential for further optimization in test-time scaling.


\section{Conclusion}
In this paper, we investigate the effectiveness and feasibility of generative ranking in large-scale industrial settings. Through both theoretical analysis and empirical results, we find that the generative architecture is the primary source of effectiveness in generative recommendation. We also introduce a novel generative architecture, named \method, which treats items as positional information and focuses on iteratively predicting user behaviors to address the inefficiencies present in existing approaches. Extensive large-scale offline and online experiments demonstrate the effectiveness and efficiency of our proposed solution.

\bibliographystyle{rusnat}
\bibliography{reference}

\end{document}